\documentclass[prl,twocolumn,showpacs,preprintnumbers,amsmath,amssymb]{revtex4}

\usepackage{graphicx}
\usepackage{bm}

\begin{document}


\title{Coherence lengths and anisotropy in MgB$_2$ superconductor}

\author{A.~Dul\v{c}i\'{c}, M.~Po\v{z}ek, D.~Paar}
\affiliation{%
Department of Physics, Faculty of Science, University of Zagreb, P. O. Box 331,
HR-10002 Zagreb, Croatia
}%

\author{Eun-Mi~Choi, Hyun-Jung~Kim, W.~N.~Kang, and Sung-Ik~Lee}
\affiliation{National Creative Research Initiative Center for Superconductivity 
and Department of Physics, \\
Pohang University of Science and Technology, Pohang 790-784,
Republic of Korea
}%


\begin{abstract}
Field and temperature microwave measurements have been carried out on MgB$_2$ thin film 
grown on Al$_2$O$_3$ substrate. The analysis reveals the mean field coherence length $\xi_{MF}$ 
 in the mixed state and a temperature independent anisotropy ratio 
$\gamma_{MF} = \xi_{MF}^{ab} / \xi_{MF}^c \approx 2$. At the superconducting transition, 
the scaling of the fluctuation conductivity yields the Ginzburg-Landau coherence length 
with a different anisotropy ratio $\gamma_{GL} =2.8$, also temperature independent. 
\end{abstract}

\maketitle

The recent discovery of superconductivity at 39~K in the simple 
binary compound MgB$_2$ \cite{Nagamatsu} has sparked a considerable effort in 
the scientific community to determine the fundamental parameters and
the nature of superconductivity in this compound. Quite surprisingly,
the usually simple determination of the upper critical field $B_{c2}$ has emerged 
as a controversial issue. The early attempts to determine $B_{c2}^c$ 
(${\bf B}\parallel c$-axis), and $B_{c2}^{ab}$ (${\bf B}\parallel ab$-plane),
have shown a large span of values and anisotropy ratio 
$\gamma=B_{c2}^{ab}/B_{c2}^c$ varying in the range 1.2-9 \cite{Buzea}. 
One could have ascribed these discrepancies to an insufficient control of the sample preparation
conditions at the early stage. However, the controversy has not been
fully settled even with the improved sample quality in the recently prepared thin
films and single crystals of MgB$_2$ \cite{Kim:02,Ferdeghini:01,Ferdeghini:02,Angst,Takahashi,
AngstComment,Welp,Zehetmayer,Lyard,Sologubenko}. One of the puzzling observations was that 
different experimental techniques often yielded
strongly diverse $B_{c2}$ values in one and the same sample. Thus, Welp et al. \cite{Welp}
have shown that resistive onset of superconductivity in a given field, which was 
traditionally taken as the upper critical field, was in disaccord with the $B_{c2}^c$
values obtained in the same sample by specific heat and magnetization measurements.
Similar discrepancy has been observed in the results of the resistive onset and
the thermal conductivity \cite{Sologubenko}. On the other hand, the onset of the diamagnetic 
response was found to corroborate with the zero resistance (or the onset of finite resistivity)
\cite{Lyard}.

The common approach in these methods is to make a choice of a percentage in cutting the 
transition curves. The corresponding points are then taken for $B_{c2}(T)$. 
Alternatively, one looks for the geometrical intersection of the tangents above and 
below the transition. None of these choices, however, is guided by a physical law describing the transition.
 
In this Letter we show that the problem of the upper critical field, and the related coherence 
length, is more subtle than implicitely assumed before. Our analysis is based on the physical 
process which defines the shape of the experimental curve, and yields unequivocally the value 
of $B_{c2}$. We find the mean field (MF) coherence length $\xi_{MF}$ as the radius of the vortex 
core in the mixed state and, separately, the Ginzburg-Landau (GL) coherence length $\xi_{GL}$ 
at the transition. The two coherence lengths are quite different in  MgB$_2$. 
The anisotropy ratios are also different ($\gamma_{GL} > \gamma_{MF}$), but both turn out to be 
temperature independent. 

	The thin film of MgB$_2$ was grown on (1$\bar{1}$02) Al$_2$O$_3$ substrate using a two-step 
method \cite{Kang,Kim:01}. Precursor thin film of B was deposited by pulsed laser deposition 
at room temperature. The B thin film was sealed together with high purity Mg into Nb tube with 
Ar atmosphere. The heat treatment was carried out at 900~$^\circ$C for 10-30 min. The film
thickness was 400 nm. X-ray diffraction indicated that the MgB$_2$ film has a highly 
$c$-axis oriented crystal structure normal to the substrate surface with no impurity phase observed.

Microwave measurements were carried out in an elliptical cavity resonating in  $_e$TE$_{111}$ 
mode at 9.3 GHz. The thin film was mounted on a sapphire sample holder and placed in the 
center of the cavity where the microwave electric field $E_{\omega}$ was maximum. 
The sample was oriented with $ab$-plane parallel to $E_{\omega}$. The measured quantities 
were the $Q$-factor of the cavity loaded with the sample and the shift of the resonant 
frequency $f$. From the complex frequency shift 
$\Delta \widetilde{\omega} / \omega = \Delta f/f  + i \Delta (1/2Q)$ one can obtain by 
inversion the complex conductivity $\widetilde{\sigma} = \sigma_1 - i \sigma_2$ of the 
film using the cavity perturbation expression \cite{Peligrad:01}.

\begin{figure}[h]
\includegraphics[width=20pc]{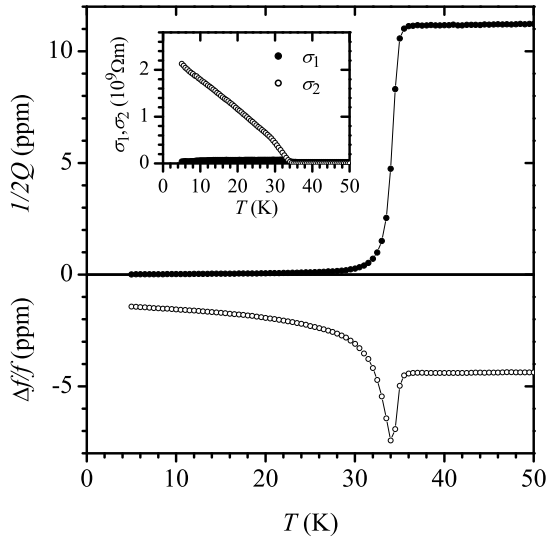}
\caption{Plots of imaginary and real parts of the complex frequency 
shift in MgB$_2$ thin film in zero magnetic field. Inset shows the corresponding conductivities.} 
\label{fig_01}
\end{figure}

Figure~\ref{fig_01} shows the experimental results in zero magnetic field. From the 
imaginary part of the conductivity $\sigma_2 = 1/\mu_0 \omega \lambda_L^2$ one can infer 
the zero temperature London penetration depth $\lambda_L(0) = 79$~nm in our film. With this 
value, and the shape of $\sigma_2$, this film is found to be between the clean and the dirty 
limit, closer to the latter \cite{Golubov:02}.

\begin{figure}[h]
\includegraphics[width=20pc]{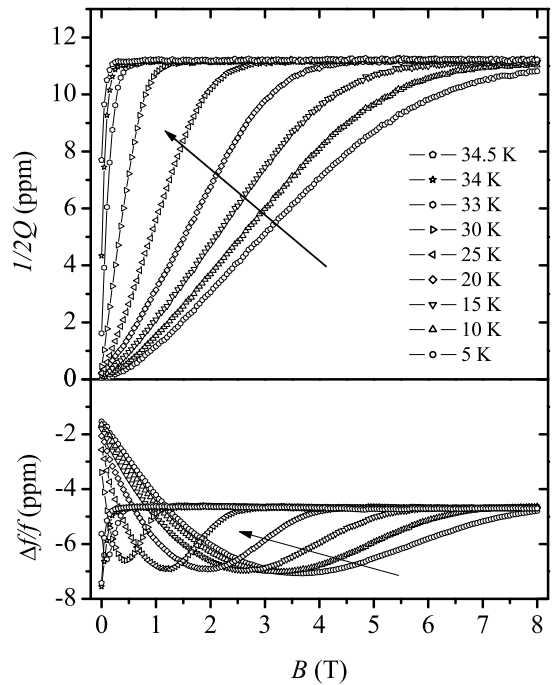}
\caption{Magnetic field dependences of the complex frequency 
shift in MgB$_2$ thin film for ${\bf B} \parallel c$. The arrows indicate increasing temperatures.} 
\label{fig_02}
\end{figure}

Here we focus on the effects of the applied magnetic field in the superconducting state. 
Figure~\ref{fig_02} shows the field dependences (${\bf B} \parallel c$) of the complex frequency shift 
at various temperatures. By inversion of these data points, one can obtain the field 
dependent complex conductivity at each given temperature. Theoretically, the response of 
the superconductor in the mixed state to an oscillating electric field $E_\omega$ is given 
by an effective complex conductivity \cite{Dulcic:93}
\begin{equation}
\frac{1}{\widetilde{\sigma}_{\mathrm{eff}}}= \frac{1-{\displaystyle \frac{B/B_{c2}}{1-i(\omega_0/\omega)}}}
{(1-{\displaystyle \frac{B}{B_{c2}}})(\sigma_1-i \sigma_2)+{\displaystyle \frac{B}{B_{c2}}} \sigma_n}+
\frac{1}{\sigma_n}\, \frac{B/B_{c2}}{1-i(\omega_0/\omega)}
\label{eq:1}
\end{equation}

The first term is due to the microwave current outside the vortex cores, and the second to 
the normal current in the cores of the oscillating vortices. The meaning of the fraction 
$B/B_{c2}$ in Eq.~(\ref{eq:1}) is the volume fraction of the sample taken by the normal vortex 
cores. The depinning frequency $\omega_0$ may change with field and temperature from 
strongly pinned case ($\omega_0  \gg \omega$) to the flux flow limit ($\omega_0 \ll \omega$). 
In Eq.~(\ref{eq:1}) the zero field conductivity is $\sigma_1 - i \sigma_2$, and $\sigma_n$ is 
the normal state conductivity. 

\begin{figure}[h]
\includegraphics[width=20pc]{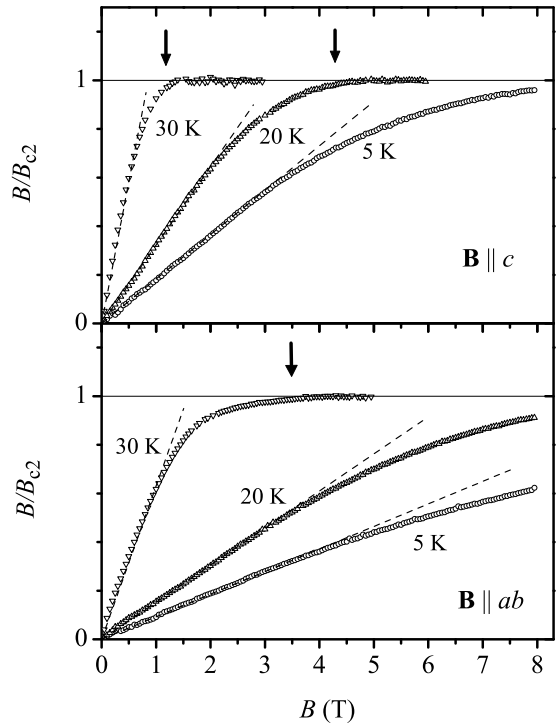}
\caption{Variations of the volume fraction of the normal electrons. The dashed lines 
mark the low field linear segments of the curves wherefrom  $B_{c2}^{MF}$ can be determined. 
The arrows indicate the $B_{c2}^{LLL}$ values obtained from the 3D LLL scaling of the 
fluctuation conductivity in Fig.~\ref{fig_04}.} 
\label{fig_03}
\end{figure}
Using the experimental field dependent complex conductivity and Eq.~(\ref{eq:1}) we have 
determined the values of $B/B_{c2}$ and $\omega_0 / \omega$. Figure~\ref{fig_03} shows some 
of the results. One observes that each of the curves has initially a constant slope (dashed 
lines in Fig.~\ref{fig_03}). It defines very precisely the value of  $B_{c2}$ at a given 
temperature. Note that in this region the actual field $B$ is much smaller than $B_{c2}$ 
so that the superconducting film is well in the mixed state. Hence, we determine, in fact, 
the mean field coherence length $\xi_{MF}$ as the radius of the normal vortex core
($B_{c2}^{MF} = \Phi_0 / 2\pi \xi_{MF}^2$, where  $\Phi_0$ is the flux quantum). 
The fundamental property of a vortex much below 
the transition to the normal state is that it contains many Landau levels as bound 
superconducting states \cite{Tinkham:75}. When the field is increased so that the transition 
to the normal state is approached, the upper Landau levels are gradually lifted and finally 
only the lowest Landau level remains. The field at which this level nucleates is 
conventionally known as the upper critical field $B_{c2}$. When the transition is very 
sharp, $B_{c2}$ can be determined straightforwardly from the single turning point. However, 
in the cases of rounded transitions, one has to consider the fluctuation conductivity and 
the scaling laws which govern the physics of the transition.

\begin{figure}[h]
\includegraphics[width=20pc]{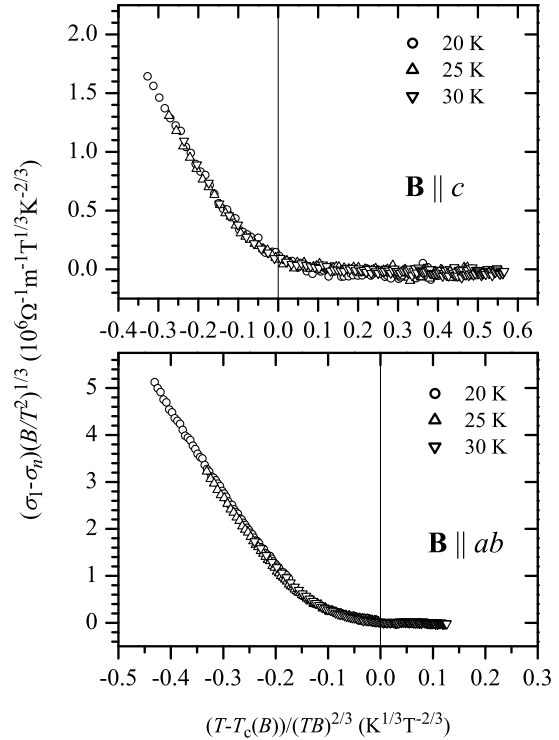}
\caption{3D LLL scaling of the fluctuation conductivity $\sigma_1-\sigma_n$.} 
\label{fig_04}
\end{figure}
In Fig.~\ref{fig_04} we show the scaling of  the fluctuation conductivity $\sigma_1 - \sigma_n$ 
according to the 3D LLL scheme \cite{Ullah:91,Ukrainczyk:95}. A very good scaling is achieved 
only with a linear choice of $T_c(B)$ line (equivalently $B_{c2}$ line) for the temperature 
interval indicated in Fig.~\ref{fig_04}. 
The corresponding values are marked by arrows in Fig.~\ref{fig_03}. It is obvious that the 
values of $B_{c2}$ obtained from 3D LLL scaling are close to the points where the normal 
state seems to be reached, but not precisely there. This feature  is due to the superconducting 
fluctuations which appear also above the mean field transition. By taking the deviation from 
the normal state behavior as the onset of superconductivity, one selects, in fact, the point 
where the superconducting fluctuations start to exhibit a contribution noticeable above the 
noise level in the experimental curve. 

\begin{figure}[h]
\includegraphics[width=20pc]{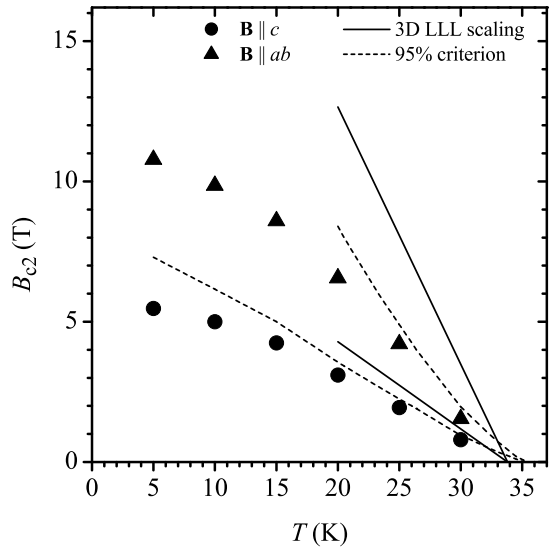}
\caption{The upper critical fields determined by various methods, $B_{c2}^{MF}$ (symbols), 
$B_{c2}^{GL}$ (full lines), and the values resulting from cutting the experimental 
curves (dashed lines).} 
\label{fig_05}
\end{figure}
The results of the present analysis are synthesized in Fig.~\ref{fig_05}. The full symbols 
represent the mean field results $B_{c2}^{MF}$ with vortices formed by a great number of 
Landau levels and having radius $\xi_{MF}$. The anisotropy ratio 
$\gamma_{MF} = \xi_{MF}^{ab} \approx  2$ is practically temperature independent. 
The full lines in Fig.~\ref{fig_05} are obtained from 3D LLL scaling. These lines delineate 
the nucleation of vortices with the lowest Landau level only. The field required for 
this nucleation is related to the Ginzburg-Landau coherence length 
$B_{c2}^{LLL} = \phi_0 / 2 \pi \xi_{GL}^2$. The anisotropy ratio is 
$\gamma_{GL}= \xi_{GL}^{ab}/\xi_{GL}^c = 2.8$, and also temperature independent in the 
interval where the 3D LLL scaling could be applied. 

We  show by the dashed lines in Fig.~\ref{fig_05} the results obtained by cutting the 
experimental curves of $1/2Q$ in Fig.~\ref{fig_02} at 95\% of the normal state value. 
The line obtained  in this way for $B_{c2}^{ab}(T)$ exhibits a positive curvature and, 
consequently, yields a temperature dependent anisotropy ratio.
Similar results have been obtained in other recent reports 
\cite{AngstComment,Welp,Zehetmayer,Lyard,Sologubenko} where various cutting 
criteria have been used rather than the scaling law.
It is not possible to 
find a single cutting level which would mimic the 3D LLL scaling procedure. It appears 
that the cutting level should be changed from one experimental curve to another at a 
different temperature in a way which is, a priori, not known. Obviously, the effects of 
the superconducting fluctuations cannot be simply accounted for by a cuting procedure.
 
One may remark that the extrapolated $B_{c2}^{LLL}$ lines in Fig.~\ref{fig_05} point to 
$T_c = 33.8$~K, while the dashed lines based on the 95\% cutting criterion reach $35.5$~K. 
With a higher percentage for the cutting level, one could reach even higher temperatures. 
These values are in the fluctuation region above the true $T_c$ defined as the temperature 
where $\xi_{GL}$ diverges.

In conclusion, we have shown that in MgB$_2$  one can distinguish the mean field coherence 
length in the mixed state, and the Ginzburg-Landau coherence length at the transition. 
The latter should be found by considering the superconducting fluctuations and the proper 
scaling law of the measured physical quantity. We find the anosotropy ratios 
$\gamma_{MF} \approx 2$ and $\gamma_{GL} = 2.8$, both with no temperature dependences.

%

\end{document}